\documentclass[]{article}

\usepackage{amssymb}
\usepackage{amsmath}
\usepackage{graphicx}
\usepackage{epsfig}
\usepackage{float}
\usepackage{subfigure}
\usepackage{psfrag}
\usepackage[T1]{fontenc}
\usepackage{color}
\usepackage{array}
\usepackage{dsfont}
\usepackage{algorithm,algorithmic}
\usepackage{tikz}
\usepackage{listings}
\usepackage{epsfig}

\usetikzlibrary{arrows,decorations.pathmorphing,backgrounds,fit,positioning,shapes.symbols,chains,shapes}
\definecolor{greenMatlab}{RGB}{46,139,87} 
\definecolor{bleuONERA}{RGB}{16,97,169}
\definecolor{grisONERA}{RGB}{64,64,66}
\providecommand{\red}[1]{\textcolor[rgb]{0.98,0.00,0.00}{#1}}

\newtheorem{remark}{Remark}
\providecommand{\lti}[0]{\textbf{LTI}~}

\providecommand{\svd}[0]{\textbf{SVD}~} %

\providecommand{\ie}[0]{\emph{i.e.}~}
\providecommand{\etc}[0]{\emph{etc.}~}
\providecommand{\eg}[0]{\emph{e.g.}~}

\newenvironment{eq}{\everymath {\displaystyle \everymath{ }} \equation}{ \endequation} %
\DeclareMathOperator*{\rank}{\mathbf{rank}}

\providecommand{\svd}[0]{\mathbf{SVD} } %
\providecommand{\pare}[1]{\left(#1\right) } %
\providecommand{\x}[0]{\mathbf{x}} %
\renewcommand{\u}{\mathbf{u}} %
\providecommand{\y}[0]{\mathbf{y}} %
\providecommand{\E}[0]{{E}} %
\providecommand{\A}[0]{{A}} %
\providecommand{\B}[0]{{B}} %
\providecommand{\C}[0]{{C}} %

\providecommand{\LL}[0]{{\mathds L}} %
\providecommand{\sLL}[0]{{\mathds L_\sigma}} %

\providecommand{\Cplx}[0]{\mathbb{C}} %
\begin{document}


\title{Interpolatory-based data-driven pulsed fluidic actuator control design and experimental validation}

\author{C. Poussot-Vassal, P. Kergus, F. Kerherv\'e, D. Sipp and
L. Cordier
\thanks{Charles Poussot-Vassal is with ONERA / DTIS, Universit\'e de Toulouse, F-31055 Toulouse, France. Pauline Kergus is with Lund University, Lund, Sweden. Franck Kerherv\'e and Laurent Cordier are with Institut Pprime CNRS - Universit\'e de Poitiers - ISAE-ENSMA - UPR 3346. Denis Sipp is with ONERA / DAAA, Meudon, France. Contact:  \texttt{charles.poussot-vassal@onera.fr}}
}
\maketitle


\begin{abstract}
Pulsed fluidic actuators play a central role in the fluid flow experimental control strategy to achieve better performances of aeronautic devices. In this paper, we demonstrate, through an experimental test bench, how the interpolatory-based Loewner Data-Driven Control (\textbf{L-DDC}) framework is an appropriate tool for accurately controlling the outflow velocity of this family of actuators. \textbf{L-DDC} combines the concept of ideal controller with the Loewner framework in a single data-driven rationale, appropriate to experimental users. The contributions of the paper are, first, to emphasise the simplicity and versatility of such a data-driven rationale in a constrained experimental setup, and second, to solve some practical fluid engineers concerns by detailing the complete workflow and key ingredients for successfully implementing a pulsed fluidic actuator controller from the data acquisition to the control implementation and validation stages.
\end{abstract}

\section{Introduction}
\label{sec:intro}

The design of active closed-loop flow controllers constitutes an important field of research in fluid mechanics. 
Active flow control is considered in many applications among which flows over open cavities (see \eg \cite{LeclercqJFM:2019}) and backward facing steps, boundary layer flows (see \eg \cite{Sipp:2016}), flows over airfoils or in combustion processes (see \eg \cite{Shaabani:2017,Willcox:2002}). The possible objectives are to maintain laminarity or delay transition to turbulence, decrease turbulence level, reduce noise, increase lift and decrease drag, enhance mixing and heat release \etc
Without detailing the specificity and the control methodology employed in each cases, in most of these contributions and the references therein, both the sensor(s) and the actuator(s) are supposed to be lumped and ideal. By this, one intends that the sensors are capable to deliver instantaneous and accurate measurements, while the actuators are able to deliver the exact control signals computed by the flow controller (\eg with no delay, no noise, continuous control signal and unbounded intervals, \etc). These developments are relevant for academic and methodological purposes. However, in order to move towards experimental applications and expect real-life validations, it is essential to consider a realistic setup instead of these idealised versions. Therefore, accurate consideration of the actuator-sensor combination is absolutely necessary. This constitutes the core of this paper.

\begin{figure}
    \centering
    \includegraphics[width=\columnwidth]{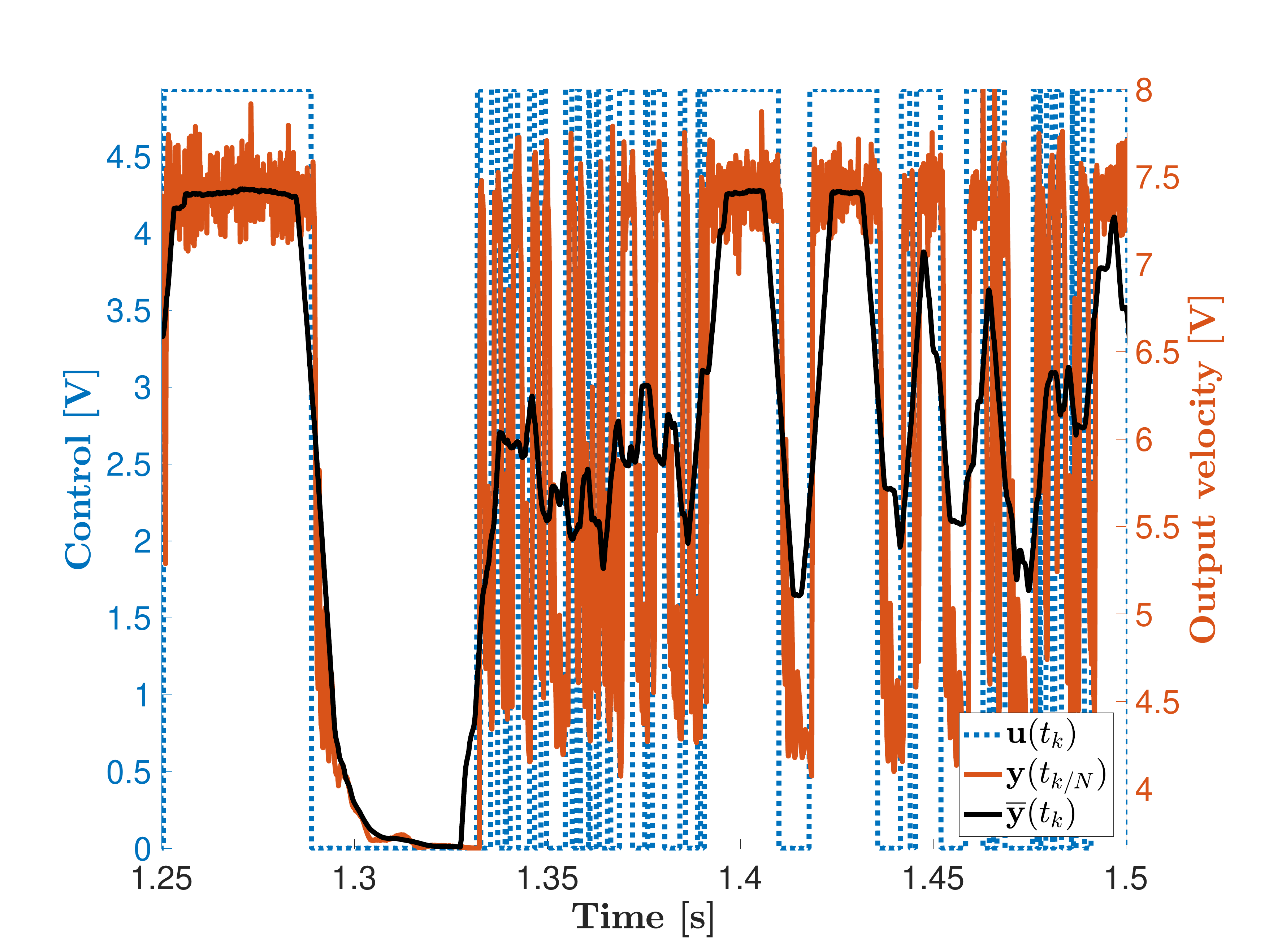}
    \caption{Extract of the open-loop data collected on the experimental setup. The output response ${\y}(t_{k/N})$ (solid orange) is obtained by feeding the \textbf{PFA} with an on/off PRBS input signal ${\u}(t_{k/N})$ (dotted blue). The averaged output $\overline{\y}(t_{k})$ (solid black) results in averaging ${\y}(t_{k/N})$ over $N$ samples. The actuator needs to be controlled to fit the reference sent by the outer-loop controller and deliver an almost continuous mean value.}
    \label{fig:OLinout}
\end{figure}

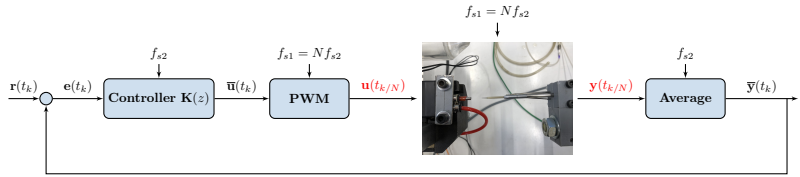
\begin{figure*}[h!]
    \centering
    \scalebox{.5}{
    
    \tikzstyle{block} = [draw, thick,fill=bleuONERA!20, rectangle, minimum height=3em, minimum width=6em,rounded corners]
\tikzstyle{block2} = [thick,rectangle, minimum height=3em, minimum width=6em,rounded corners]
\tikzstyle{sum} = [draw, thick,fill=bleuONERA!20, circle, node distance=1cm]
\tikzstyle{input} = [coordinate]
\tikzstyle{output} = [coordinate]
\tikzstyle{pinstyle} = [pin edge={to-,thick,black}]
\tikzstyle{connector} = [->,thick]

\begin{tikzpicture}[auto, node distance=2cm,>=latex']
    \node [input, name=input] {};
    \node [sum, right of=input] (sum) {};
    \node [block, right of=sum, node distance=3cm] (controller) {\textbf{Controller} $\mathbf K(z)$};
    \node [pinstyle, above of=controller, node distance=1.2cm] (samplingK) {$f_{s2}$};
    \draw [connector] (samplingK) -- node[name=hK] {} (controller.90);
    \node [block, right of=controller, node distance=4cm] (pwm) {\textbf{PWM}};
    \node [pinstyle, above of=pwm, node distance=1.2cm] (samplingPWM) {$f_{s1}=N f_{s2}$};
    \draw [connector] (samplingPWM) -- node[name=hPW] {} (pwm.90);
    \node [block2, right of=pwm, node distance=5cm] (system) {\includegraphics[width=4cm]{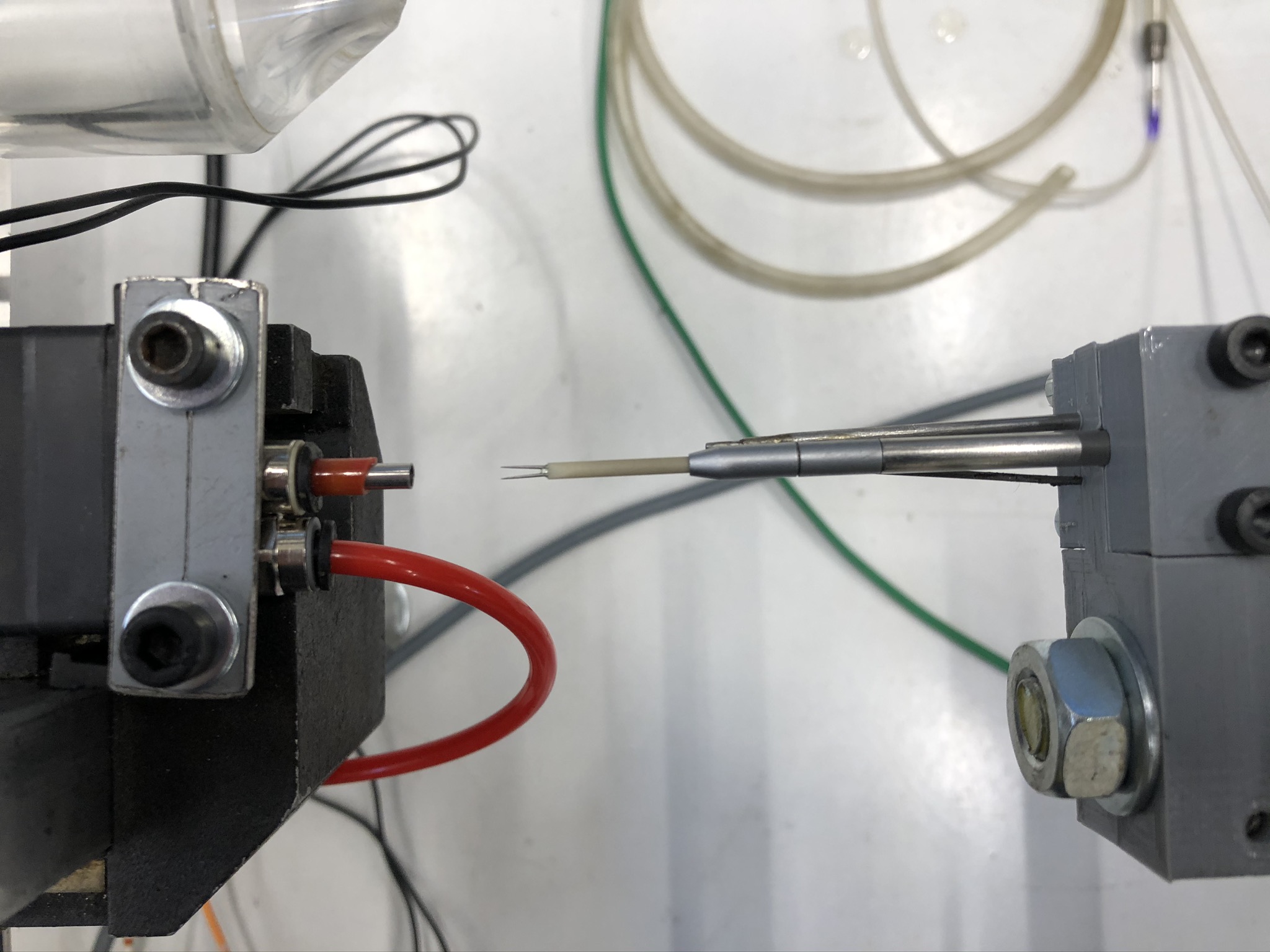}};
    \node [pinstyle, above of=system, node distance=2.3cm] (samplingSystem) {$f_{s1}=N f_{s2}$};
    \draw [connector] (samplingSystem) -- node[name=hPW] {} (system.90);
    \node [block, right of=system, node distance=5cm] (mean) {\textbf{Average}};
    \node [pinstyle, above of=mean, node distance=1.2cm] (samplingMean) {$f_{s2}$};
    \draw [connector] (samplingMean) -- node[name=hhPW] {} (mean.90);
    \draw [connector] (controller) -- node[name=u] {${\overline \u(t_k)}$} (pwm);
    \draw [connector] (system) -- node[name=u] {$\red{\y(t_{k/N})}$} (mean);
    \node [output, right of=mean, node distance=3cm] (outputMean) {};

    \draw [connector] (input) -- node {$\mathbf r(t_k)$} (sum);
    \draw [connector] (sum) -- node {$\mathbf e(t_k)$} (controller);
    \draw [connector] (pwm) -- node {\red{$\mathbf u(t_{k/N})$}} (system);
    \draw [connector] (mean) -- node {$\overline \y(t_{k})$} (outputMean);
    \draw [connector] (outputMean)+(-0.3cm,0) -- ++(-0.3cm,-2cm) -| node [near start] {} (sum.south);
\end{tikzpicture}
    
    }
    \caption{Overview of the inner-loop considered. \textbf{Controller} $\mathbf K(z)$ is the control law to be computed (sampled at frequency $f_{s2}$), Pulsed Width Modulation (\textbf{PWM}) block transforms the continuous signal into on/off values (sampled at frequency $f_{s1}$) and \textbf{Average} block is a down-sampling  function  providing the mean value of the input signal. The system is illustrated by its top view photo, where left side represents the \textbf{PFA} and right side, the Pressure Sensor (\textbf{PS}).}
    \label{fig:closedLoopScheme}
\end{figure*}

Among the different flow actuator technologies available, which may be classified as mechanical (\eg surfaces using electrical, hydraulic, morphing actuation) or fluidic (\eg pulsed or continuous blowing, synthetic jet), the pulsed fluidic actuator (\textbf{PFA}) stands as a simple, ergonomic, economic and affordable solution. The latter is therefore widely used in fluid mechanics for control application. The most widely used \textbf{PFA} type are valves which control the output mass flow rate / pressure issued from a reservoir at rest with fixed pressure and temperature. A considerable amount of studies can be found in the literature on the attempts, with more or less success, to control flows in various applications. While their response time and operating frequency can be high enough compared to the main characteristic time of the flow to manipulate (at least one order of magnitude larger), it is noteworthy that this class of actuators is mainly used to excite the flow within the receptivity range of its most unstable modes, namely the large-scale structures. Their level of authority is however restricted since the \textbf{PFA}, alone, acts on the flow on a limited spatial domain. In practical applications, these actuators are therefore usually installed in arrays to provide enough energy to gain authority over the considered flow region and mechanism to be controlled. Beside this, one major drawback is that they work either completely open or closed and are thus on/off systems, see Figure \ref{fig:OLinout}. To be integrated in a global flow control scheme, these actuators may be equipped with sensors measuring the current outflow velocity and be accurately controlled at high frequency with an inner controller so that the velocity of the blown air follows the reference control signal provided by the outer controller. Indeed, these actuators are not ideal as usually assumed, presenting asymmetric dynamics and noise, as shown on Figure \ref{fig:OLinout}. On the basis of the above comments, there is a need for providing practitioners and applied fluid engineers a systematic and simple approach to design a controller tailored to \textbf{PFA} devices, allowing to track a given reference signal fed by an outer flow controller. As mentioned previously, in practice multiple \textbf{PFA} are generally employed in parallel. Due to manufacturing and installation versatility, some discrepancies in the characteristic responses are most of the time observed between the \textbf{PFA}s installed. One additional need is therefore to  provide a methodology that can be easily applied on multiple installed devices \cite{Ternoy:2013}.

The main objective of the present contribution is therefore to provide practitioners and experimental researchers a simple but yet effective methodology and practical workflow to implement pulsed fluid actuators flow controllers directly from experiments. This is achieved through the  \textbf{L-DDC} algorithm (Loewner Data-Driven Control). This technique relies on the definition of an ideal controller, derived from a reference model \cite{Campi:2002,FormentinIET:2012}, which allows to use the Loewner framework \cite{Mayo:2007} to construct a reduced order controller, simple to implement. 
While the contribution is more methodological rather than theoretical, authors believe that the interpolatory-based data-driven control design process presented in this paper solves practical problems faced by practitioners, namely, the complete \textbf{PFA} controller design using open-loop data collected directly on the experimental setup only. This approach has proven to be effective on infinite dimensional systems \cite{KergusLCSS:2020}, for numerical control \cite{VuilleminWC:2020} and relates to data-driven stability analysis \cite{PoussotGTA:2020}.
%

The major benefit of this data-driven rationale is that the control design only requires one single set of open-loop data collected on the actuator, allowing to account for variations and discrepancies from an actuator to an other and thus to design a control law tailored to each system. Instead of spending time and energy in an identification and control design process, the data-driven approach is less costly to deploy in practice and tailored to each actuator
. In addition, as the proposed workflow does not require any optimisation iteration, its application remains easy in an experimental context.  Finally, given a set of open-loop data and an objective closed-loop specification, the proposed process automatically finds the controller structure and gains.

The control and experimental setup and closed-loop architecture are described in Section \ref{sec:setup}. Section \ref{sec:method} recalls the interpolatory-driven \textbf{L-DDC} design approach, and details the key steps to follow for proper implementation. Section \ref{sec:result} is devoted to the illustration and analysis of the obtained experimental results. Conclusions and comments finally close the paper in Section \ref{sec:conclusion}.


\section{Experimental setup and closed-loop description}
\label{sec:setup}

\subsection{Overall control setup}

A schematic view of the inner control loop is given in Figure \ref{fig:closedLoopScheme}. From a control engineer perspective, the flow control may be referred to as the outer-loop (out of this paper scope) while the pulsed fluidic actuator controller is referred to as the inner-loop. Following Figure \ref{fig:closedLoopScheme}, we are interested in the inner-loop only and more specifically in the control design, implementation and validation on a real \textbf{PFA}. With reference to Figure  \ref{fig:closedLoopScheme}, $\mathbf r(t_k)$ denotes the reference signal (typically the control signal provided by an outer controller) at time $t_k$,  $\overline\y(t_k)$ the measurement signal at time $t_k$ (typically the instantaneous velocity measured at the output of the \textbf{PFA}), $\mathbf e(t_k)=\mathbf r(t_k)-\overline\y(t_k)$ the error signal  at time $t_k$. Then $\overline{\u}(t_k)$ is the reference control signal computed at time $t_k$ by the sampled-time controller $\mathbf K(z)$ to be designed. Then,  $\u(t_{k/N})$ is the on/off control signal, at time $t_{k/N}$, modulated in pulse width by the Pulsed Width Modulation (\textbf{PWM}) block. The picture in Figure \ref{fig:closedLoopScheme} represents (left) the top view of the \textbf{PFA} and (right) the Pressure Sensor (\textbf{PS}), sensing $\y(t_{k/N})$, here velocity, at time $t_{k/N}$, which is then down-sampled and averaged at time $t_{k}$, leading to $\overline{\y}(t_{k})$. As suggested on Figure \ref{fig:closedLoopScheme}, two sampling frequencies denoted as $f_{s2}$ and $f_{s1}=Nf_{s2}$ ($N\in\mathbb{N}$), referring to sampling times $t_{k/N}$ and $t_k$ respectively, are required to implement the controller. Details of this multi-sampling setup is given later in this section.

\subsection{Experimental setup}

The experimental test bench is composed of a \textbf{PFA} and a hot-wire probe (\textbf{PS}) as shown on the photo on Figure \ref{fig:closedLoopScheme}. The acquisition and generation of the command laws are carried out thanks to a real-time hardware system (National Instruments NI PXI-1042 mainframe).

The \textbf{PFA} used in the present study is a SX11F-BH micro-valve from SMC which can operate, according to the manufacturer, up to 1 kHz at a mean flow rate of 50 L/min. The asymmetric response times at opening and closing, again according to the manufacturer, are $0.55$ms and $0.4$ms respectively. For the present purpose, the valve is alimented with compressed air at constant pressure ($2.2$bar). A short tube of $5$mm diameter is used at the valve exit. The instantaneous velocity at 3D downstream the tube exit is surveyed thanks to a hot-wire probe (Dantec 55P11) connected to a 55M10 DISA constant temperature anemometer (CTA). Note that since the purpose of the present paper is to demonstrate the potential of the methodology introduced, no calibration of the hot-wire probe is effected. The voltage of the hot-wire probe is therefore directly considered as representative of the exit velocity and used as the output signal $\y(t_{k/N})$. From an application point of view, \textbf{PFA} equipped with outflow velocity sensors need to be developed (here an external hot-wire probe system is used).  This is currently being investigated in the Micro Opto Electro Mechanical Systems (MEMS) community. 




\subsection{Pulsed width modulation and average blocks}

As the \textbf{PFA} works with only on/off control, a \textbf{PWM} function has naturally been implemented on the software side of the setup to transform the sampled-time signal $\overline\u(t_k)$ provided by the controller $\mathbf K(z)$ into $\u(t_{k/N})$, a stair signal with varying length (or duty cycle $D$). In addition, in order to take into account for noise measurement, since the probe sensor is not ideal, an average block is introduced and which operates on the measurement signal. The following presents the \textbf{PWM} and \textbf{Average} blocks in details.
\bigskip

\subsubsection{PWM block} 

 The \textbf{PWM} block uses a rectangular impulsion signal taking values between $\u_{\text{min}}$ and $\u_{\text{max}}$ and which length is modulated. This modulation results in variation of the mean of the signal $\overline\u(t_k)$ to convert. If one considers an impulsion with a high frequency $f_{s1}$ and a duty cycle $D\in[0,1]$, the averaged value $\u^{\text{avg}}(t_{k})$ of the resulting signal is given by
\begin{equation}
\begin{array}{rcl}
     \u^{\text{avg}}(t_{k}) &=& \dfrac{1}{T_{s2}} \displaystyle\int_0^{T_{s2}} \overline\u(t_k) dt_k \\
     &=& \dfrac{1}{T_{s2}} \bigg( \displaystyle\int_0^{DT_{s2}} \u_{\text{max}} dt_k + \displaystyle\int_{DT_{s1}}^{T_{s2}} \u_{\text{min}} dt_k \bigg) \\
     &=& D \u_{\text{max}} + (1-D)\u_{\text{min}}, 
\end{array}
\label{eq:pwm}
\end{equation}
where $T_{s1}=1/f_{s1}$ and $T_{s2}=1/f_{s2}$. Obviously, the \textbf{PWM} should be sampled at rate $N$ times higher than that of the signal $\u(t_k)$ to be modulated (with $N\in\mathbb N$). In practical applications, a simple way to generate the \textbf{PWM} is to use the intersection method which simply requires a saw-tooth carrier signal denoted $\u_c(t_k)$, with frequency $f_{s2}$ and amplitude from $\u_{\text{min}}=\min \overline\u(t_k)$ to $\u_{\text{max}}=\max \overline\u(t_k)$, that should be compared to the incoming signal $\overline\u(t_k)$. When $\u_c(t_k)>\overline\u(t_k)$, then $\u(t_{k/N})=\u_{\text{max}}$, and $\u(t_{k/N})=\u_{\text{min}}$ otherwise. Note that in the present case, the carrier signal has the same frequency as the control signal $\overline\u(t_k)$.
\smallskip

\subsubsection{Average block} 

The average block consists in averaging the measurement value $\y(t_{k/N})$ over $N$ past samples. The resulting output is a down-sampled signal $\overline\y(t_k)$, which represents the averaged output to be controlled and that should track $\mathbf r(t_k)$. The main purpose of such block is to partially filter the measurement noise, making the averaged value $\overline\y(t_k)$ more representative of the output state than the instantaneous value $\y(t_{k/N})$.

\begin{remark}[About $N$] \label{rem:N}
For some applications, the parameter $N$ is dictated by the hardware and setup. However, most of the time, it can be chosen by the user and is thus an additional tuning parameter. A rule of thumb is to chose $N\geq 10$ to ensure that the \textbf{PWM} block will be able to translate the control signal into an accurate binary output. Still, large $N$ leads to too strong filtering of the measured signal and may result in irrelevant behaviour. On the other side, low $N$ may lead to noisy data and thus inaccurate data-driven controller design. Consequently, authors advice users to use this parameter accordingly to the setup limitations and use it as a trade-off.
\end{remark}

\subsection{Signals characteristics and specifications}

Now the experimental setup and \textbf{PWM}/\textbf{Average} blocks have been presented, let us summarise the signals characteristics as follows:
\begin{itemize}
    \item $\mathbf r(t_k)$ (sampled at $f_{s2}$), is the reference signal to be tracked. This signal is continuous and is fed by the outer control. For the considered future application of turbulence control, its bandwidth is below $f_c=5$Hz\footnote{Note that this bandwidth may be considered as way too slow for fluid engineer. Still, this does not affect the method and higher bandwidth will be consiered in future works.} 
    The knowledge of this signal characteristics is essential in the construction of the objective performance $\mathbf{M_{f_j}}(s)$ (see Section \ref{sec:method}).
    \item $\y(t_{k/N}) \in [\y_{\text{min}}~\y_{\text{max}}]$ (sampled at $f_{s1}$), is the measurement signal obtained by the \textbf{PS}. It allows measuring the air flow velocity at the exit of the actuation. Note that the distance of the probe from the \textbf{PFA} exit has an impact on the control law by adding a delay in the time response of the device. This is an additional reason justifying for the data-driven approach: the design can be reproduced easily for different distances or/and different probe sensors. 
    \item $\overline\y(t_k) \in [\y_{\text{min}}~\y_{\text{max}}]$ (sampled at $f_{s1}$), is the averaged and down-sampled value of $\y(t_{k/N})$ on blocks of duration $1/f_{s1}$ .
    \item $\overline \u(t_k)\in [0~\u_{\text{max}}]$ (sampled at $f_{s2}$), is the control signal provided by the \textbf{PFA} controller to be modulated by the \textbf{PWM} block. The sampling frequency of this signal determines the carrier and frequency of the  \textbf{PWM}.
    \item $\u(t_{k/N})\in\{0,\u_{\text{max}}\}$ (sampled at $f_{s1}$), is the effective modulated (pulsed) control signal sent to the \textbf{PFA}. 
\end{itemize}

In our configuration, the controller $\mathbf K(z)$ and the \textbf{PWM} modules respectively run at frequency $f_{s2}=100$Hz and $f_{s1}=1000$Hz (thus $N=10$ is the \textbf{PWM} multiplicity factor). Now the setup and signals characteristics have been introduced, let us describe the main result, namely the workflow adopted to obtain the controller $\mathbf K(z)$  structure and gains.

The method proposed in this paper, detailed in the next section, aims at designing a controller $\mathbf K(z)$ to improve the tracking performances of the couple \textbf{PFA}-\textbf{PS}. Indeed, as this system is stable and minimum-phase, the system does not need to be stabilised. The emphasis will therefore be put on enhancing closed-loop performances while maintaining stability.


\section{Main result: data-driven fluidic actuator control design}
\label{sec:method}

\subsection{Overview of the approach and control objective}

Considering the closed-loop structure of Figure \ref{fig:closedLoopScheme}, the objective is to track a reference signal $\mathbf r(t_k)$ \textcolor{black}{up to a frequency of $f_c=5$Hz, being the maximal frequency of the outer loop (not detailed here)}. To this aim and for the simplicity and flexibility of the implementation, a frequency-domain interpolation-based data-driven approach has been chosen. The employed workflow is summed up in Algorithm \ref{workflow} and key steps are described in the remaining of this section.
\begin{algorithm}
\caption{\textbf{L-DDC} based \textbf{PFA} control design} \label{workflow}
\begin{algorithmic}[1]
\REQUIRE A desired reference closed-loop model $\mathbf{M_{f_j}}(s)$, a \textbf{PWM} frequency $f_{s1}$ and ratio $N\in\mathbb N$.
\ENSURE $\mathbf K(z)$ sampled at $f_{s2}$, leading to similar performances to $\mathbf{M_{f_j}}(s)$ when inserted in the loop.
\STATE (Section \ref{subsec:openloop}) Collect open-loop data
\begin{itemize}
\item  Generate an exciting signal using a pseudo-random binary sequence denoted $\u_{\text{prbs}}(t_{k/N})$.
\item Excite the \textbf{PFA} system with $\u(t_{k/N})=\u_{\text{prbs}}(t_{k/N})$, collect $\y(t_{k/N})=\y_{\text{prbs}}(t_{k/N})$ and compute the mean values leading to the so-called open-loop data 
$$\{t_k,\overline \u_{\text{prbs}}(t_k),\overline{\y}_{\text{prbs}}(t_k)\}_{k=1}^M.$$
\item Compute the corresponding Fourier signals $\tilde\u_{\text{prbs}}(\imath\omega_i)$ and $\tilde\y_{\text{prbs}}(\imath\omega_i)$, leading to the frequency-domain experimental transfer data as 
$$
\bigg\{\imath\omega_i,\mathbf{\Phi}_i=\frac{\tilde\y_{\text{prbs}}(\imath\omega_i)}{\tilde\u_{\text{prbs}}(\imath\omega_i)}\bigg\}_{i=1}^N.
$$
\end{itemize}
\STATE (Section \ref{subsec:lddc}) Apply \textbf{L-DDC}
\begin{itemize}
\item Compute the ideal controller $\mathbf K^\star(\imath\omega_i)$.
\item Use interpolation-based data-driven model construction and obtain the full order interpolant controller $\mathbf K_{\text{full}}(s)$.
\item Apply rank revealing factorisation for controller order reduction leading to $\mathbf{\tilde K}(s)$.
\end{itemize}
\STATE (Section \ref{subsec:lddc}) Compute $\mathbf K(z)$, the discrete-time version $\mathbf {\tilde K}(s)$ using any discretisation method, with frequency sampling $f_{s2}=f_{s1}/N$.
\end{algorithmic}
\end{algorithm}


\subsection{Open-loop data acquisition and frequency response construction}
\label{subsec:openloop}

The first three steps of Algorithm \ref{workflow} are now detailed and commented. These steps consists in exciting the system to obtain a relevant open-loop frequency response. 

\subsubsection{Construction of a PRBS exciting signal}

Open-loop data can be obtained using a pseudo random binary sequence signal as input of the system\footnote{Note that many other exciting signals exist but this one suits well to such a fluidic system.}. This signal consists in an on/off sequence of varying length. A "random" sequence of these signals, whose duration should last enough to reach steady state output and short enough to excite frequencies above the cut-off one of the system, is defined. The Fourier transform of the sequence $\u_{\text{prbs}}$ should be constant over the frequencies until a cut-off and is thus equivalent to a white noise over these frequencies. In practice, due to the rectangular shape of each pulse given as
\begin{equation}
    \u_{\text{prbs}}(t_{k/N}) =  \textbf{rect}_{T_{s2}}(t/T_{s2}) \left\{
    \begin{array}{ll}
    1 & \text{, $\forall t \in [-T_{s2}/2,T_{s2}/2]$} \\
    0 & \text{, elsewhere}
    \end{array}
    \right.
\end{equation}
the corresponding Fourier transform reads
\begin{equation}
    \tilde \u_{\text{prbs}}(f) = \dfrac{\sin(\pi f T_{s2})}{\pi f} = T_{s2} \mathbf{sinc}(\pi T_{s2} f),
\end{equation}
where $T_{s2}=1/f_{s2}$. Thus, harmonics can be observed every $k\pi T_{s2}$ (for $k\in \mathbb N$) and should be treated in the post-processing phase.

\subsubsection{Frequency transfer construction}

Based on the input $\u_{\text{prbs}}(t_{k/N})$ and corresponding output $\y_{\text{prbs}}(t_{k/N})$ signals, one computes the averaged signals $\overline\u_{\text{prbs}}(t_{k})$ and  $\overline\y_{\text{prbs}}(t_{k})$ and  their Fourier transform $\tilde \u_{\text{prbs}}(\imath\omega_i)$ and $\tilde \y_{\text{prbs}}(\imath\omega_i)$ respectively ($\imath=\sqrt{-1}$). Then, the cross correlation transfer of these signals are used to construct $\mathbf \Phi_i$ as 
\begin{equation}
    \mathbf \Phi_i(\imath\omega_i) = \dfrac{\tilde \y_{\text{prbs}}(\imath\omega_i)}{\tilde \u_{\text{prbs}}(\imath\omega_i)}. 
\end{equation}
In opposition to the spectral energy density, the cross (or inter-correlated) spectral density is a complex number which gain represents the interaction power and which arguments represents the phase between $\overline \u_{\text{prbs}}(t_k)$ and $\overline \y_{\text{prbs}}(t_k)$. Now the open-loop data have been collected and treated, the Loewner Data-Driven Control (\textbf{L-DDC}) process is deployed \cite{KergusLCSS:2019,KergusLCSS:2020}.

\subsection{Loewner Data-Driven Control}
\label{subsec:lddc}
The \textbf{L-DDC} algorithm allows to design a controller on the basis of frequency-domain data from the system to be controlled. As presented in \cite{KergusPhD:2019,KergusLCSS:2019}, the \textbf{L-DDC} approach covers the choice of the reference model, the definition of the ideal controller (step 2, bullet 1) and its interpolation (step 2, bullet 2) and reduction based on the Loewner framework  (step 2, bullet 3), but also a data-driven stability analysis of the resulting closed-loop. These different aspects of the \textbf{L-DDC} control design are detailed hereafter for the proposed application.

\begin{remark}[Continuous vs. sampled-time \textbf{L-DDC}]
Note that, in the end, a continuous controller $\mathbf{\tilde K}(s)$ is obtained and then discretised to obtain the control law $\mathbf{K}(z)$. An hybrid version of the \textbf{L-DDC}, allowing to obtain directly a discrete controller for a continuous system, also exists and can be found in \cite{VuilleminWC:2020}, but has not been used here for practical reasons (remove work and work decoupling).
\end{remark}

\paragraph{Reference model specification}
\label{ref_model}

As detailed in \cite{KergusLCSS:2020}, the choice of the specifications is a key aspect of the \textbf{L-DDC} procedure, and can be a difficult task for a practitioner. Indeed, in the general case, the user should take into account the performance limitations of the system to be controlled to avoid compensations of instabilities between the system and the controller, which would lead to an internally unstable closed-loop. The performance limitations of a system are determined by its own instabilities, imposing interpolatory conditions on the closed-loop to be achievable. When the system is only known through data, its instabilities, and the associated performance limitations, must be found in a data-driven way, as it is proposed in \cite{KergusLCSS:2020}, using the model-free stability analysis introduced in \cite{cooman2018model} and \cite{cooman2018estimating}.

In the present case, as pointed in in Section \ref{sec:setup}, the actuator-sensor couple is stable and minimum-phase. This result can also be determined from the experimental data as in \cite{cooman2018estimating}, or using the Loewner-based stable projection as explained in \cite{PoussotGTA:2020} and chap. 5 of \cite{PoussotGTA:2020}. Therefore any stable transfer function can be used as reference model. In this favorable case, the \textbf{L-DDC} procedure is greatly simplified: there is no instability to estimate and any stable transfer function can be used as a reference model. To that extent, the reference model can be chosen as a simple, well-known and desirable model such as a first-order one given as
\begin{equation}
    \mathbf{M_{f_j}}(s)=\frac{1}{\frac{s}{2\pi f_j}+1},
    \label{reference_model}
\end{equation}
with a unitary static gain for precision purposes (no steady-state error). The only parameter left to choose is the desired closed-loop speed, determined by the frequency $f_j\in\mathbb R_+$: it is to be chosen according to the bandwidth frequency of the phenomenon to be controlled in the outer-loop. In this study, three frequencies are considered, $f_1=1$Hz, $f_3=3$Hz and $f_5=5$Hz, corresponding to reference models $\mathbf{M_{f_1}}$, $\mathbf{M_{f_3}}$ and $\mathbf{M_{f_5}}$, leading to the design of three $\mathbf K(z)$ discrete-time controllers, respectively denoted $\mathbf{K_{f_1}}(z)$, $\mathbf{K_{f_3}}(z)$ and $\mathbf{K_{f_5}}(z)$. In what follows, the design is detailed for $\mathbf{M}=\mathbf{M_{f_1}}$ only.

\paragraph{Ideal controller definition}

Once the reference model $\mathbf{M}$ is fixed, it is possible to use the available open-loop data $\{\imath\omega_i,\mathbf{\Phi}_i\}_{i=1}^N$ to define the ideal controller $\mathbf{K}^\star$. It is the unique \lti controller that would have given the desired performances during the experiment if inserted in the closed-loop. By definition, samples of its frequency response can be computed as follows, $\forall i=1\dots N$:
\begin{equation}
    \mathbf{K}^\star(\imath\omega_i) = \mathbf{\Phi}_i^{-1} \mathbf{M}(\imath\omega_i) (I - \mathbf{M}(\imath\omega_i))^{-1},
    \label{ideal_controller}
\end{equation}
where $I$ denotes the identity matrix of appropriate dimension.

Thanks to the ideal controller, the identification can be shifted from the system to the controller, and this is what makes the \textbf{L-DDC} procedure data-driven (see \eg \cite{Campi:2002}). Indeed, the main step, detailed in the next paragraph, consists in obtaining a (rational) realisation $\mathbf{K}_{\text{full}}$ of the ideal controller $\mathbf{K}^\star$ that satisfies the following interpolatory conditions, $\forall i=1\dots N$:
\begin{equation}
    \mathbf{K}_{\text{full}}(\imath\omega_i) = \mathbf{K}^\star(\imath\omega_i), 
    \label{CI}
\end{equation}
and then to derive $\mathbf{\tilde K}$, a reduced-order version of it.

\paragraph{Full and reduced controller design via Loewner}

The Loewner framework allows to find a rational \lti model $\mathbf{K}_{\text{full}}$ achieving \eqref{CI}. Recent descriptions of the Loewner realisation landmark by Mayo and Antoulas are available in  \cite{Mayo:2007}. In short, let the interpolation points $\imath\omega_k$ be divided in two equal subsets as follows ($\lambda_i\in\Cplx$ and $\mu_j\in\Cplx$):
\begin{equation}
\{z_k\}_{k=1}^{2m} =\{\mu_j\}_{j=1}^{m} \cup \{\lambda_i\}_{i=1}^{m}=\{\imath \omega_k\}_{k=1}^{N}.
\label{eq:shift}
\end{equation}
The Loewner realisation framework by \cite{Mayo:2007}  consists in building the \emph{Loewner} $\LL \in \Cplx^{m\times m}$ and \emph{shifted Loewner} $\sLL \in \Cplx^{m\times m}$ matrices defined as follows (in the single input single output case), for $i=1,\dots,m$ and $j=1,\dots,m$:
\begin{eq}
\begin{array}{rcl}
[\LL]_{j,i} &=& \dfrac{\mathbf{K}^\star(\mu_j) - \mathbf{K}^\star(\lambda_i) }{\mu_j - \lambda_i} \\
\,[\sLL]_{j,i} &=& \dfrac{ \mu_j\mathbf{K}^\star(\mu_j) - \lambda_i\mathbf{K}^\star(\lambda_i) }{\mu_j - \lambda_i}
\end{array}.
\label{eq:loewnerMatrices}
\end{eq}
Then, the model $\mathbf{K}^m$ given by the following descriptor realisation,
\begin{equation}
\E^m  \delta\left\{\x(\cdot)\right\} = \A^m \x(\cdot) + \B^m \u(\cdot)\text{ and }
\y(\cdot) =\C^m \x(\cdot),
\label{eq:loewnerDescrR}
\end{equation}
where $\E^m = -\LL$, $\A^m = -\sLL$, $[\B^m]_k = \mathbf{K}^\star(\mu_k)$ and $[\C^m]_k = \mathbf{K}^\star(\lambda_k)$ (for $k=1,\ldots,m)$ and whose related transfer function 
\begin{equation}
\mathbf K^m(\xi) = \C^m\pare{\mathbf \xi \E^m-\A^m}^{-1}\B^m,
\label{eq:loewnerDescrTF}
\end{equation}
satisfies \eqref{CI}. In \eqref{eq:loewnerDescrR}, $\delta(\cdot)$ denotes the derivative operator in the continuous-time and the forward one in the sampled-time. Similarly, $\xi$ is the Laplace variable in the continuous case and the $z$ operator in the sampled case. Assuming that the number $2m=N$ of available data is large enough, then  it has been shown in \cite{Mayo:2007} that a minimal model of dimension $n < m$ that still interpolates the data can be built with a projection of \eqref{eq:loewnerDescrR} provided that, for $k=1,\ldots,2m$, $\rank\pare{z_k \LL - \sLL} = \rank\pare{ [\LL,\sLL]} = \rank\pare{[\mathbb{L}^H,\sLL^H]^H} = n$. In that case, let us denote by $Y \in \Cplx^{m \times n}$ the matrix containing the first $n$ left singular vectors of $[\LL,\sLL]$ and $X \in \Cplx^{m \times n}$  the matrix containing the first $n$ right singular vectors of $[\LL^H,\sLL^H]^H$. Then, $\mathbf{K}_{\text{full}}:(Y^H \E^m X ,Y^H \A^m X ,Y^H \B^m,\C^m X)$ is the minimal McMillian degree rational function also satisfying  the interpolation conditions \eqref{CI} (see details in \cite{AntoulasChapSpringer:2010}).

However, $\mathbf{K}_{\text{full}}$ is often of very high order and such a controller would be too complex to be implemented. In the present case, the interpolation gives a high-order minimal realisation, $n=290$, principally due to  noisy data.

\begin{figure}
    \centering
    \includegraphics[width=\columnwidth]{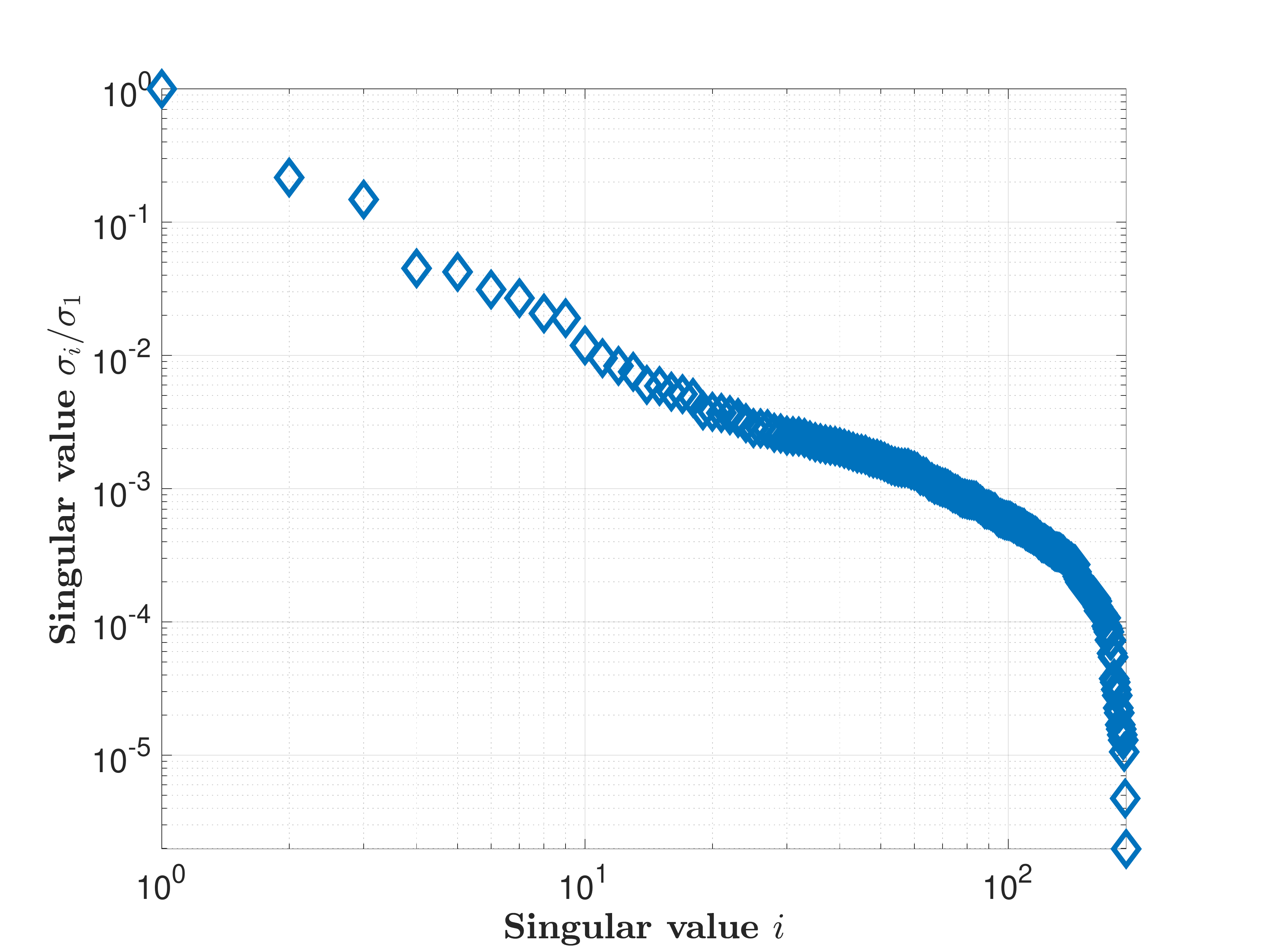}
    \caption{Singular Value Decomposition of the Loewner pencil: one mode is sufficient to describe the dynamic of the ideal controller.}
    \label{fig:svdK}
\end{figure}

Similarly to the rank truncation performed above, the Loewner framework allows to control the complexity of the identified controller. It is possible to obtain a $r$-th order reduced controller $\mathbf{\tilde K}$ of the minimal realisation $\mathbf{K}_{\text{full}}$ by applying the \textbf{SVD} on the Loewner pencil and truncating with a user defined order $r<n$. The order of the controller becomes a parameter that can be tuned by the user, accordingly to the Loewner singular value decay. In the present case, the \svd decay of the associated Loewner pencil is given on Figure \ref{fig:svdK} (see \cite{AntoulasSurvey:2016} for practical details and chap. 2 of \cite{PoussotHDR:2019} for some applications).

\begin{figure}
    \centering
    \includegraphics[width=\columnwidth]{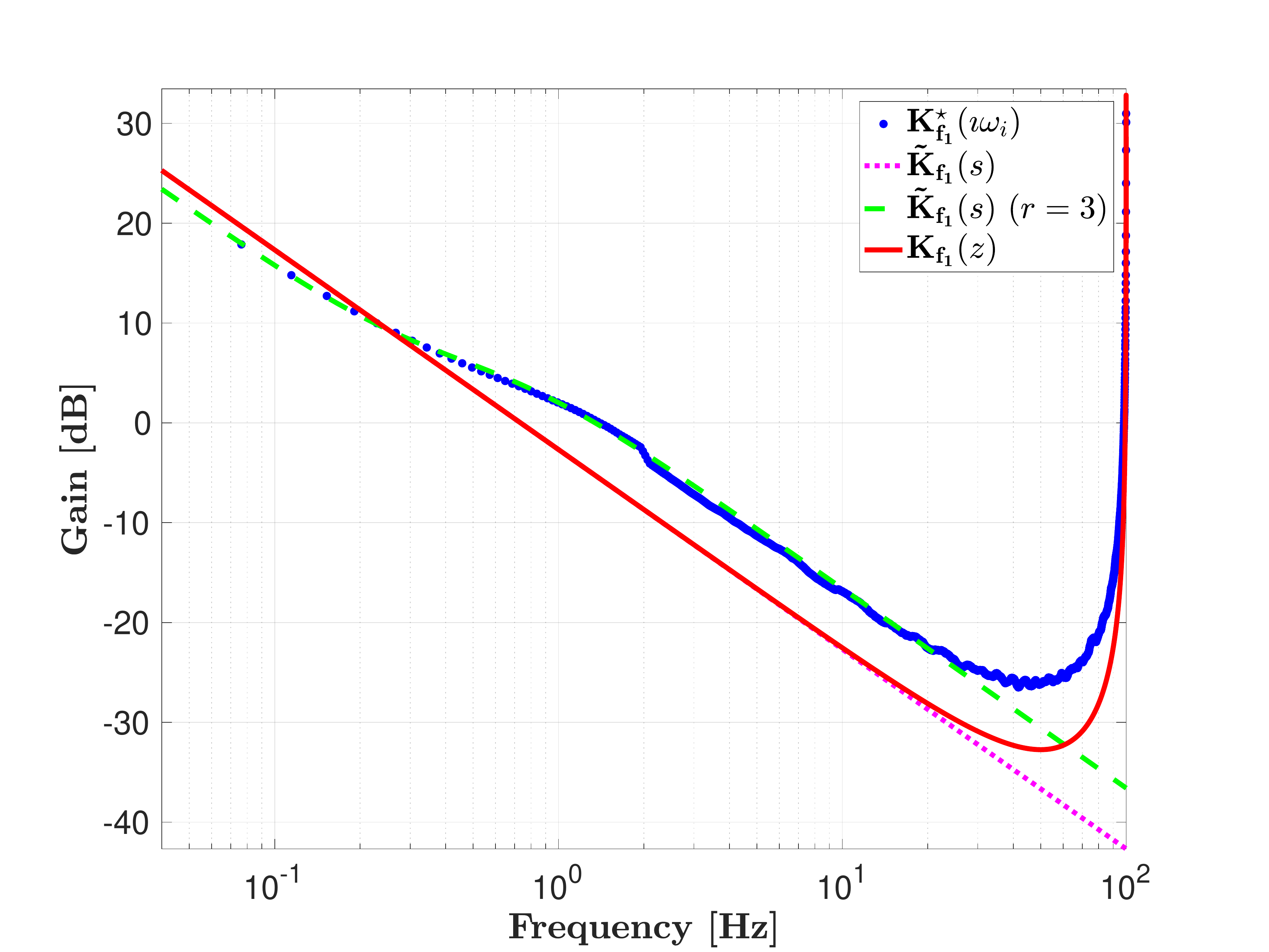}
    \caption{Gain of the frequency responses of the ideal controller $\mathbf K^\star$ evaluated at the available frequencies (blue dots), of the estimated controller $\mathbf{\tilde K_{f_1}}(s)$ of order $r=1$ (dotted magenta), $r=3$ (dashed green) and its time-sampled version $\mathbf{K_{f_1}}(z)$ (solid red).}
    \label{fig:K}
\end{figure}

Figure \ref{fig:svdK} suggests an order $r=3$ as it provides almost 80\% singular value contribution (\ie $\sigma_1+\sigma_2+\sigma_3 \approx 0.8\sum_{i=1}^n\sigma_i$). However selecting an order $r=1$ still achieves to 50\% of the information $\sigma_1 \approx 0.5\sum_{i=1}^n\sigma_{i}$. Such choice leads to a pure integral control action, which discrete-time model reads
\begin{equation}
\mathbf K(z) = \frac{k_j}{z-1},
\label{eq:K}
\end{equation}
where $k_j\in\mathbb R_+$ is the integral gain, computed by the proposed \textbf{L-DDC} procedure. Figure \ref{fig:K} shows the responses of the ideal controller, as well as the reduced integral controllers obtained with $r=1$ (continuous and discrete), and compare it to the order $r=3$ one. Note that the control order remains a parameter that the designer can choose according to its implementation limitations. Here the simple integral choice has been considered as a good trade-off between complexity and performances. 

\paragraph{Preliminary verification}

The reduction of the controller model implies that the  closed-loop, denoted $\mathbf{T_{K}}$, obtained when inserting the reduced and sampled-time $\mathbf{K}$ controller, will not be equal to the desired reference model $\mathbf{M}$. Therefore, it is interesting to compute the frequency response of the expected closed-loop $\mathbf{T_{K}}$ on the basis of the available open-loop data $\{\imath\omega_i,\mathbf{\Phi}_i\}_{i=1}^N$ from the system as follows, $\forall i=1\dots N$
\begin{equation}
     \mathbf{T_{K}}(\imath \omega_i)=(I+\mathbf{\Phi}_i\mathbf{ K}(e^{\imath \omega_iT_{s2}}))^{-1}\mathbf{\Phi}_i\mathbf{ K}(e^{\imath \omega_iT_{s2}}).
    \label{cl_estimate}
\end{equation}

The result of the expected closed-loops for three different controllers is illustrated later on Figure \ref{fig:CL_freq}. Note that for $r=3$, authors considered that the gain was not important enough with respect to experimental benefits of such a simple integral structure, and is therefore not illustrated.

\paragraph{About stability}

The choice of the reference model presented earlier ensures that the ideal controller stabilises the system internally. However, there is no stability guarantee regarding the use of a reduced-order controller. In \cite{KergusLCSS:2020}, the small-gain theorem is derived in a data-driven way: it is shown that limiting the controller modelling error during the reduction step ensures internal stability. This stability test is very conservative and another solution would be to use the projection-based technique of \cite{cooman2018model} or the one in \cite{PoussotHDR:2019} to conclude regarding the stability of the closed-loop. Both approaches are data-driven one and may be used to guess the stability. Note also that in the single input-single output case, the Nyquist criteria can also be considered.


\section{Experimental results and analysis}
\label{sec:result}
\subsection{Considered configurations}

Based on the presented control design rationale and preliminary validation procedures presented in Section \ref{sec:method}, we are now ready to implement and validate the control setup on the real experimental test bench. The real-time acquisition and control functions, including the control law $\mathbf{K}(z)$, the \textbf{PWM} and \textbf{Average} blocks as schematised on Figure \ref{fig:closedLoopScheme}, are implemented within a LabView interface. For the considered flow system, three different controllers denoted $\mathbf{K_{f_j}}(z)$ with the same structure as in \eqref{eq:K}, sampled at $f_{s2}=100$Hz, corresponding to three different performance objectives $\mathbf{M_{f_j}}(s)$ as defined in \eqref{reference_model}, for $j=\{1,3,5\}$, are constructed. These may be read as,
\begin{itemize}
    \item $\mathbf{K_{f_1}}$, where $k_1 = 0.0462$ ($\mathbf{M_{f_1}}$ with $f_1=1$Hz),
    \item $\mathbf{K_{f_3}}$, where $k_3 = 0.1385$ ($\mathbf{M_{f_2}}$ with $f_3=3$Hz), and
    \item $\mathbf{K_{f_5}}$, where $k_5 = 0.2309$ ($\mathbf{M_{f_3}}$ with $f_5=5$Hz).
\end{itemize}
For the considered application, following Figure \ref{fig:closedLoopScheme} notations, $N=10$ and thus $f_{s1}=1000$Hz were chosen. The closed-loop performances are first validated with a sine sweep signal and second, with a more realistic reference trajectory.

\subsection{Closed-loop sine sweep reference signal}

\begin{figure}
    \centering
    \includegraphics[width=\columnwidth]{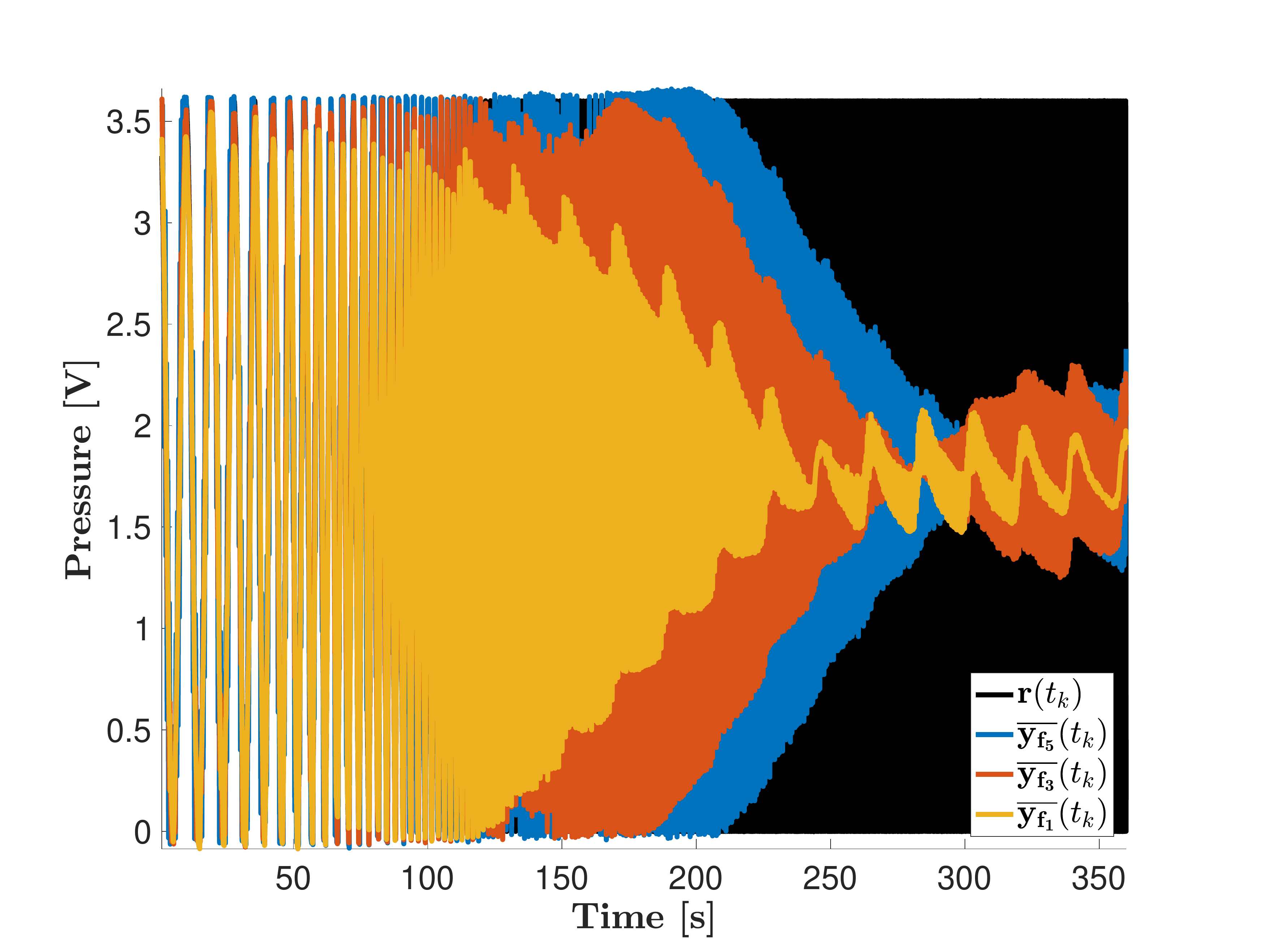}
    \caption{Closed-loop time-domain response to a sine sweep in frequency as reference. Coloured shade, the different mean output values $\overline{\mathbf{y_{f_j}}}(t_k)$ (corresponding to $\overline{\mathbf{y}}(t_k)$ on Figure \ref{fig:closedLoopScheme}) for different $\mathbf{K_{f_j}}$ controllers (corresponding to $\mathbf{K}$ on Figure \ref{fig:closedLoopScheme}).}
    \label{fig:CL_time}
\end{figure}

We first apply a reference signal $\mathbf r(t_k)$ being a frequency sweep from $0.1$Hz to $30$Hz, of duration $1000$s and amplitude ranging from the minimal to the maximal possible values of the system. The mean values $\overline{\mathbf{y_{f_j}}}(t_k)$\footnote{Signal $\overline{\mathbf{y_{f_j}}}(t_k)$ refers to $\overline{\mathbf{y}}(t_k)$ on Figure \ref{fig:closedLoopScheme}, for the $j$-th configuration ($j=\{1,2,3\}$).} for all three configurations and time-domain responses are first reported on Figure \ref{fig:CL_time}, in response to a sweep reference signal $\mathbf r(t_k)$. 

\begin{figure}
    \centering
    \includegraphics[width=\columnwidth]{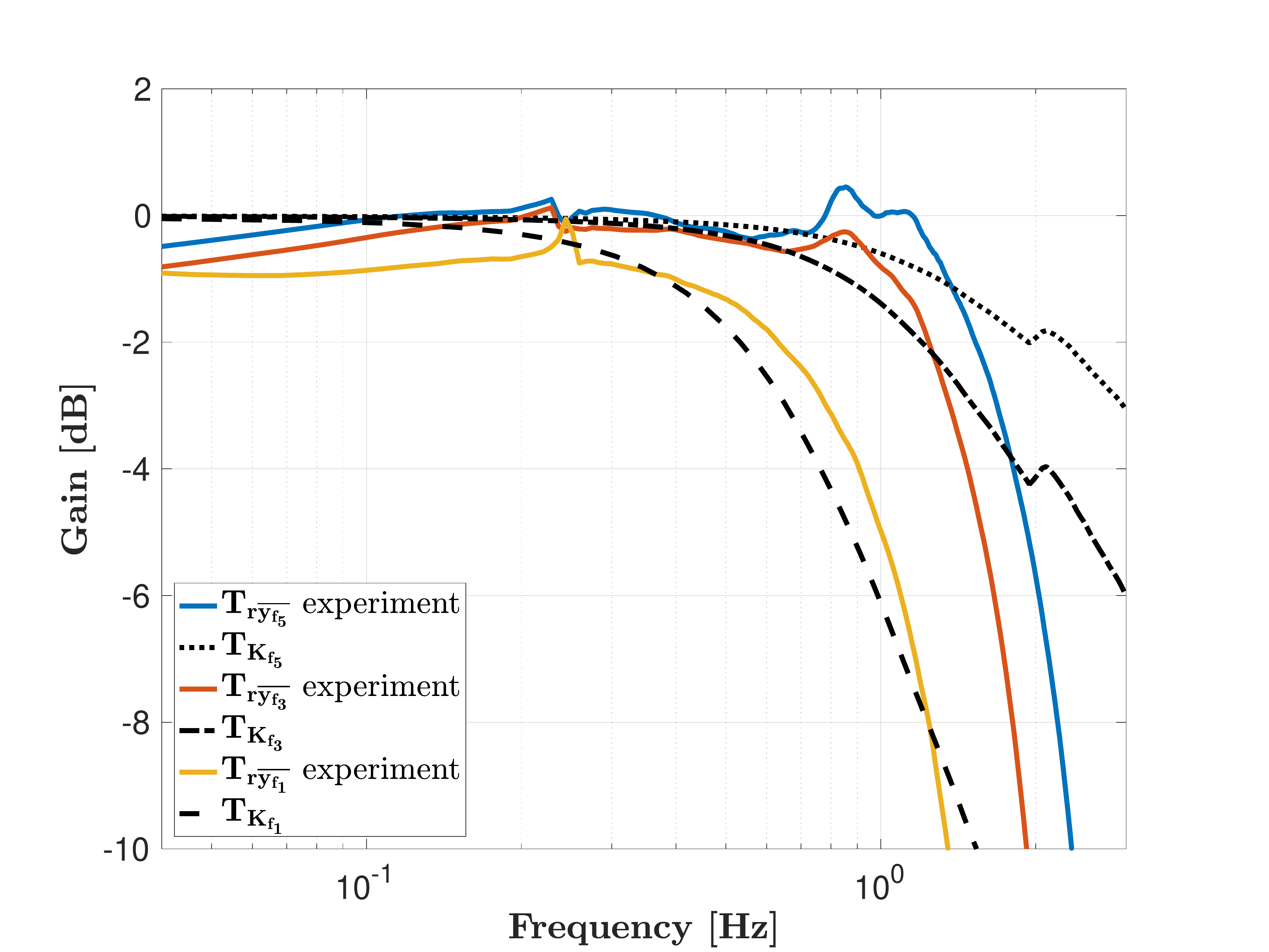}\\
    \includegraphics[width=\columnwidth]{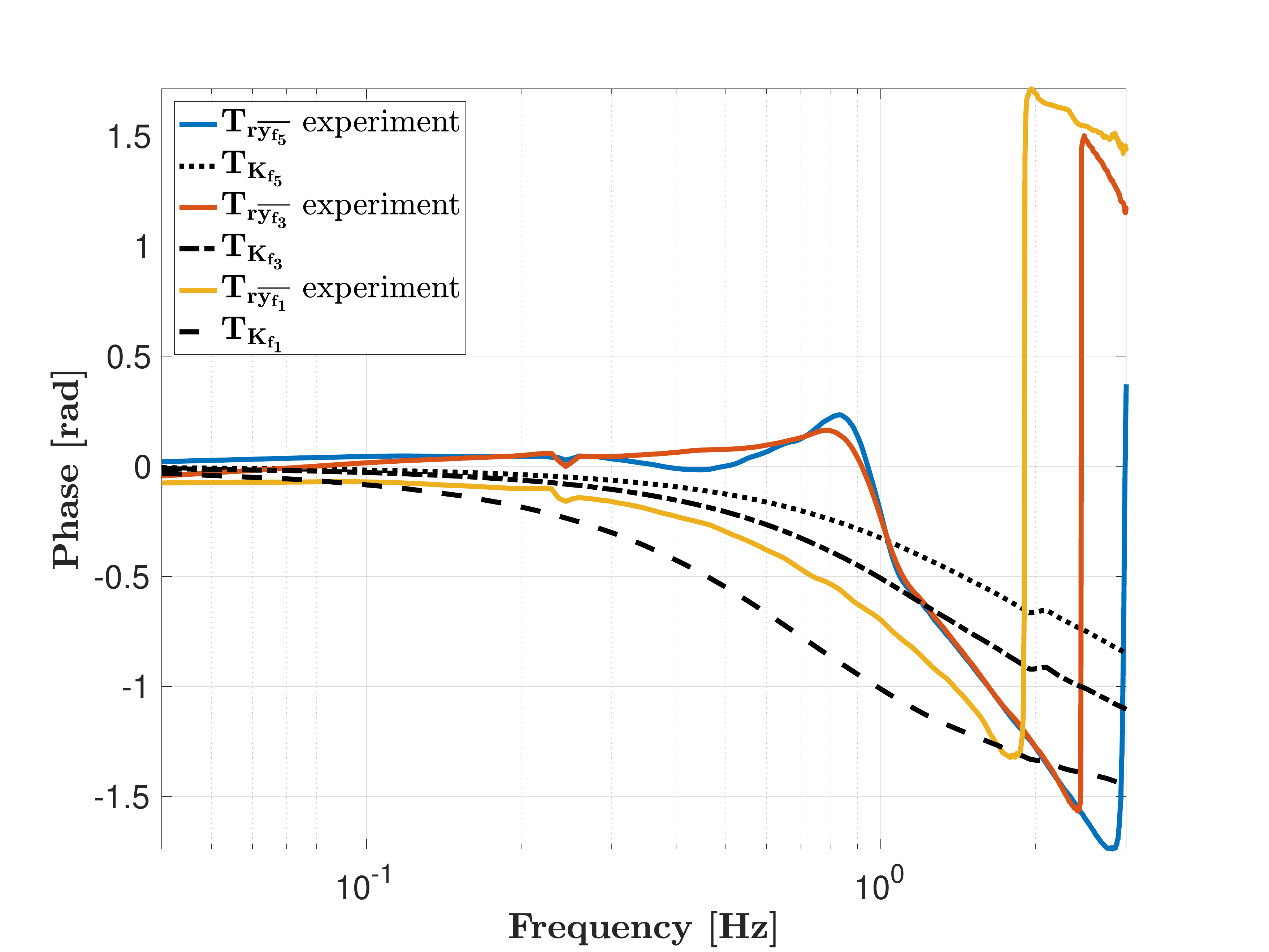}
    \caption{Comparison of the gain (top) and phase (bottom) of the closed-loop frequency-domain responses of $\mathbf{T_{r\overline{y_{f_j}}}}$ (from $\mathbf r$ to $\overline{\mathbf{y_{f_j}}}$), obtained with the designed controllers $\mathbf{K_{f_j}}$, compared to the expected closed-loops $\mathbf{T_{K_{f_j}}}$.}
    \label{fig:CL_freq}
\end{figure}

\begin{figure*}
    \centering
    \includegraphics[width=1\columnwidth]{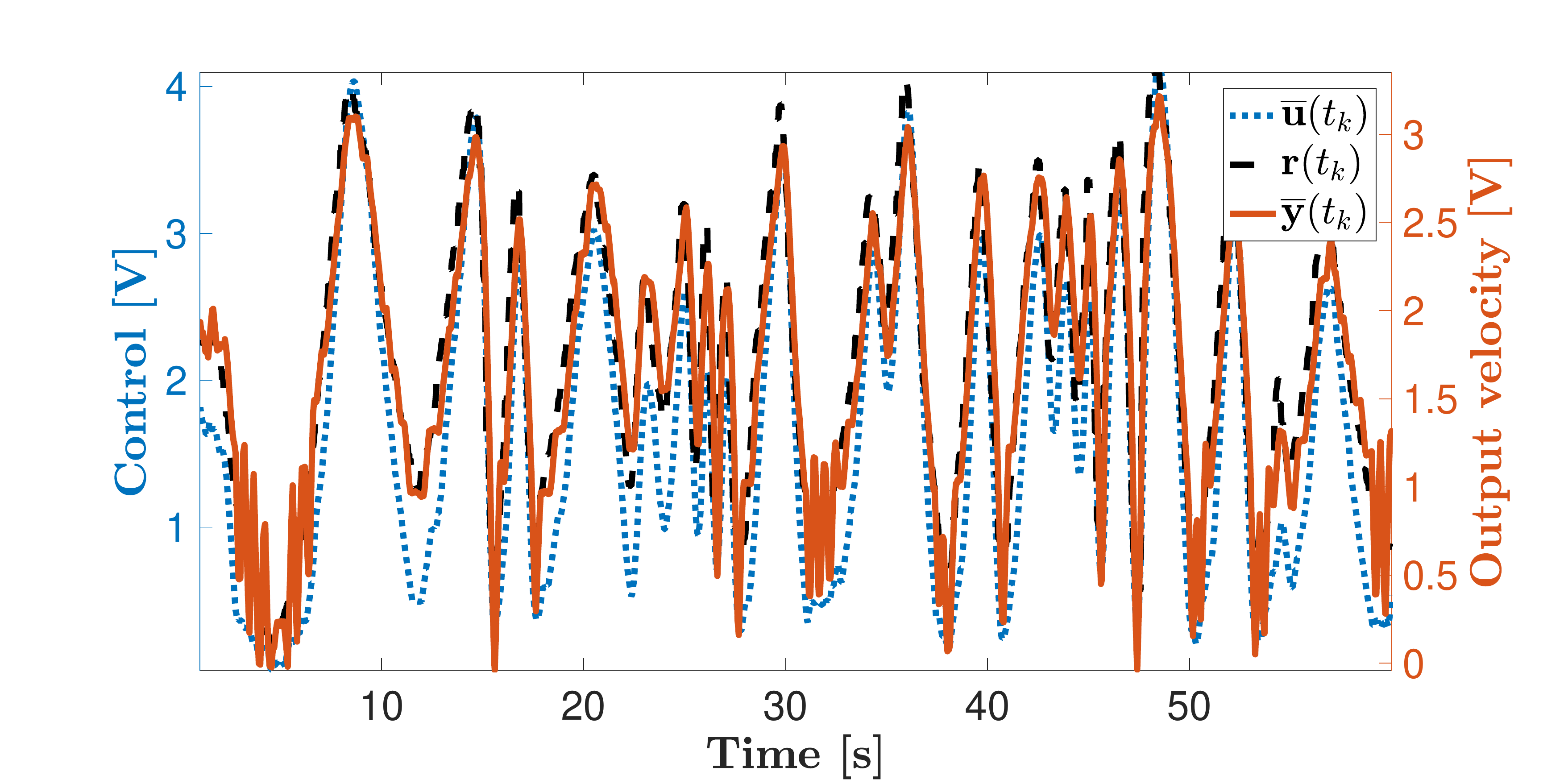}
    \caption{Closed-loop response using controller $\mathbf{K_{f_1}}$, of the averaged output $\overline{\y}(t_k)$ (solid orange) to a variable reference trajectory $\mathbf r(t_k)$ (dashed black). Averaged control signal $\overline{\u}(t_k)$ (dotted blue).}
    \label{fig:CL_tracking}
\end{figure*}

The averaged output flow velocity well tracks the reference signal considered. In all three cases, the low frequencies are well tracked, which is already an important result for \textbf{PFA} systems and in view of fluid flow control. One can also observe that the tracking is more accurate in high frequencies with the more aggressive controller $\mathbf{K_{f_5}}$, synthesised to target a bandwidth of $f_5=5$Hz, validating the reference model approach of the data-driven procedure. 
Figure \ref{fig:CL_freq} compares the experimental frequency-domain responses from $\mathbf r$ to $\overline{\mathbf{y_{f_j}}}$, denoted $\mathbf{T_{r\overline{y_{f_j}}}}$, with $\mathbf{T_{K_{f_j}}}$, the expected closed-loop computed in \eqref{cl_estimate}. For each controllers, the resulting closed-loop $\mathbf{T_{r\overline{y_{f_j}}}}$ provides a similar frequency response trends as the one expected and illustrated with $\mathbf{T_{K_{f_j}}}$. Some differences persist but can fairly be attributed to the nonlinear and non-symmetric nature of the pulsed  actuator. Indeed, as illustrated in Figure \ref{fig:OLinout} and  shown later in the section, when a reference amplitude close to zero is requested, some stick-like behaviour are observed, leading to a loss of accuracy. Still,  the obtained control law designed solely through the lens of a pure data-driven approach, provides the required performances.

\subsection{Closed-loop using realistic reference signals}

Let us demonstrate the efficiency of the proposed control with a more realistic reference signal $\mathbf r(t_k)$. Figure \ref{fig:CL_tracking}  shows the tracking performance of the mean output $\overline{\y}(t_k)$ and control $\overline{\u}(t_k)$ signals using the first controller $\mathbf{K_{f_1}}$\footnote{On Figure \ref{fig:CL_tracking}, indices are removed and notations are the one of Figure \ref{fig:closedLoopScheme}.}. Clearly, Figure \ref{fig:CL_tracking} exposes really satisfactory results in terms of tracking. As previously pointed, one can remark that some chattering artefact and difficulties appear when tracking in low amplitude references. This observation can be correlated to the asymmetric actuator characteristic which produces a stick-like behaviour in the valve opening. Even though not in the scope of this study, one way to limit this effect is to add high frequency noise in the control signal to avoid the problem, but at the price of an actuator fatigue (note that the approach is similar to solutions used in friction control). The averaged control signal $\overline{\u}(t_k)$ is also reported. Interestingly, a simple integral controller action, obtained from a single open-loop data collection, using only on/off signals, allows tracking a complex reference signals with a good accuracy. 


\section{Conclusions and discussions}
\label{sec:conclusion}

One underlying objective of this paper was to bridge the gap between fluid experts and control engineers by providing  practitioners a simple way to adjust a control law for a pulsed (on/off) fluidic actuator, without spending too much energy in a time consuming identification-control process. To this aim, a frequency-domain data-driven approach (celebrated as \textbf{L-DDC}) is used. Such a procedure, which solely relies on a single experimental data set, allows to find the order and controller gains tailored to the system under consideration. The complete approach has been applied and validated through an experimental setup. As an interesting result, a simple integral action was shown to be enough for ensuring such a tracking task. Authors believe that the proposed workflow presents a valid alternative to the complex identify and control approach, especially in the experimental wind tunnel context where experiments are expensive and time is limited.  In the coming steps of the project, 96 similar pulsed fluidic actuators and their associated (integral) control will be installed and used over a complete one meter wing span. These 96 actuators may  be lumped as a single one with a given bandwidth in order to control a flow phenomena over a given geometry, through an outer-loop control law. Connections with positive systems are also under investigation to integrate actuators limitations \cite{BriatSIAM:2020}\footnote{Supplementary video material is also available on the first author's webpage, https://sites.google.com/site/charlespoussotvassal/visual.}.


\end{document}